\documentclass{article}
\usepackage{spconf,amsmath,amssymb,graphicx}

\usepackage{enumitem}
\setlist{nosep, leftmargin=14pt}

\usepackage{todonotes}
\usepackage{orcidlink,textcomp}
\usepackage[capitalise]{cleveref}

\hypersetup{
    colorlinks=true,
    }

\def\x{{\mathbf x}}
\def\X{{\mathcal X}}
\def\xhat{{\mathbf{\hat x}}}
\def\y{{\mathbf y}}
\def\A{{\mathbf A}}
\def\M{{\mathbf M}}

\def\Tg{{\mathbb{T}_g}}
\def\Tgi{{\mathbb{T}_{g^{-1}}}}
\def\L{{\mathcal{L}}}
\def\GT{{\text{GT}}}
\def\ft{{f_\theta}}

\title{Fully unsupervised dynamic MRI reconstruction via diffeo-temporal equivariance}
\name{Andrew Wang\orcidlink{0000-0003-0838-7986} and Mike Davies\orcidlink{0000-0003-2327-236X}}
\address{Institute for Imaging, Data and Communications, School of Engineering, University of Edinburgh}

\begin{document}
%
\maketitle
\begin{abstract}
Reconstructing dynamic MRI image sequences from undersampled accelerated measurements is crucial for faster and higher spatiotemporal resolution real-time imaging of cardiac motion, free breathing motion and many other applications. Classical paradigms, such as gated cine MRI, assume periodicity, disallowing imaging of true motion. Supervised deep learning methods are fundamentally flawed as, in dynamic imaging, ground truth fully-sampled videos are impossible to truly obtain. We propose an unsupervised framework to learn to reconstruct dynamic MRI sequences from undersampled measurements alone by leveraging natural geometric spatiotemporal equivariances of MRI. \textbf{D}ynamic \textbf{D}iffeomorphic \textbf{E}quivariant \textbf{I}maging (\textbf{DDEI}) significantly outperforms state-of-the-art unsupervised methods such as SSDU on highly accelerated dynamic cardiac imaging. Our method is agnostic to the underlying neural network architecture and can be used to adapt the latest models and post-processing approaches. Our code and video demos are at \href{https://github.com/Andrewwango/ddei}{this https URL}.
\end{abstract}


\section{Introduction}

Dynamic MRI reconstruction is crucial to noninvasively capture physiological processes that are only visible in real-time video, such as heart motion \cite{lyu_state---art_2024}, free-breathing motion \cite{liu_rare_2020}, vocal-tract imaging \cite{lim_3d_2019} and blood-flow but often results in artifact-contaminated images. This is due to acquisition of heavily undersampled k-t-space measurements, resulting in a highly ill-posed inverse problem.

Classical methods such as cine imaging assume that the cardiac cycle repeats with each heartbeat, allowing images to be acquired over multiple beats (\textit{gating}) but ignores any aperiodic dynamics. Current compressed sensing (CS) methods require hand-crafted priors, lengthy and expensive inference-time optimisation and acceleration rates may be limited \cite{lyu_state---art_2024}.

Deep learning (DL) offers promising results. However, it is not only difficult, but \textit{impossible} to obtain ground truth (GT) sequences since it is impossible to capture fully-sampled images at the same sufficient frame rate. Therefore, all proposed supervised models, such as transformer and diffusion models \cite{lyu_state---art_2024}, use fully-sampled pseudo-GT obtained from classical methods such as gating. However, this constitutes a \textit{data crime} \cite{shimron_implicit_2022} - models trained on such data will never be able to learn true physiological motion and its irregularities, which are often of interest in medical imaging.

Therefore, in order to make dynamic MR images faster and cheaper to obtain, cleaner, and able to image true motion, unsupervised methods are required that can learn to image from noisy undersampled raw non-gated measurements alone, particularly since these are easily collected in the wild.

\subsection{Contributions}

\begin{enumerate}
    \item A new unsupervised framework for undersampled video reconstruction applied to dynamic MRI (also easily applicable to other dynamic medical imaging scenarios);
    \item An extension to equivariant imaging (EI) that simultaneously exploits temporal and diffeomorphic invariances;
    \item Demonstration of state-of-the-art unsupervised imaging performance for dynamic MR imaging of cardiac motion.
\end{enumerate}

\section{Method}

\subsection{Dynamic MRI background}

Let $\y=\{\y_t\}_{t=1}^{t=T}\in\mathbb{C}^{H_k\times W_k\times T}$ be a sequence of time-binned single-coil complex k-t-space samples, where each $\y_t$ is of a set of $\tau$ k-t-space samples, e.g. lines in Cartesian sampling or radial spokes, where $\tau$ is chosen such that the frame-rate is high enough that the motion is sufficiently captured. Let $\x^{(i)}=\{\x_t\}_{t=1}^{t=T}\in\mathbb{C}^{H\times W\times T}$ be the $i$th unknown GT image sequence that we wish to estimate from a dataset $i\in I$. Then, with suitable vectorisation, these are related by

\begin{equation}
    \y^{(i)}_t=\A^{(i)}_t\x^{(i)}_t + \mathbf{\epsilon}, \A^{(i)}_t=\M^{(i)}_t \mathbf{F}
    \label{eq:forward}
\end{equation}

\noindent where $\M=\{\M_t\}_{t=1}^{t=T}\sim\mathcal{M}$ is a time-varying mask representing the locations of the k-t-space samples (depending on masking strategy) randomly drawn from a distribution, $\mathbf{F}$ is the 2D Fourier operator and $\mathbf{\epsilon}$ is additive i.i.d noise. Note this framework easily generalises to multi-coil (with appropriate inclusion of coil sensitivities) and 3D+t imaging scenarios.

In the ill-posed inverse problem, the goal is to train a neural network (NN) $\ft$ to reconstruct $\xhat=\ft(\y)$ (or $\ft(\y,\A)$ when knowledge of $\A$ is to be included in the network), but the number of k-t-space measurements $m=H_kW_kT$ is much smaller than the size of the images $n=HWT$ ($(i)$ omitted for brevity). The supervised method trains with a loss $\L_\text{sup}=\L(\ft(\y),\x_\GT)$ which requires GT, where $\L$ is any common loss such as the L2, L1 or perceptual losses. The naive unsupervised method trains a baseline using measurement consistency (MC) $\L_\text{MC}=\L(\A \ft(\y),\y)$ which cannot recover any information from the nullspace of $\A$ ($\x:\A\x=\mathbf{0}$).

\subsection{Diffeo-temporal group equivariance}

\begin{figure}[tb]
  \centering
  \includegraphics[width=0.48\textwidth]{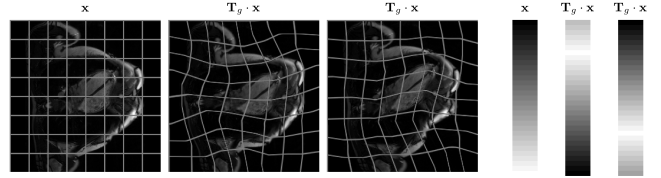}
  \caption{Demo of our group transforms. Note these are just visualisations: we never have access to the GT images $\x$. Left: sample MR image and two random diffeomorphisms $\in\mathcal{D}$. Right: demo x-t sequence and two random time-shifts $\in\mathcal{T}$.}
  \label{fig:transforms}
\end{figure}

We posit that the unknown \textit{set} of all MRI 2D+t image sequences $\X$ is invariant to geometric spatiotemporal transformations defined by a group $G$ i.e. $g\cdot x\in\X\;\forall g\in G,x\in\X$. We define the temporal group as time-shifts and reflections i.e. the dihedral group of order T $\mathcal{T}=\text{Dih}_T$. Since we have soft deformable tissue \cite{gan_deformation-compensated_2022}, the set should also be invariant to the $C^1$-diffeomorphism group $\mathcal{D}$. Finally, we take the direct product of the above $G=\mathcal{T}\bigotimes\mathcal{D}$ and construct the group action $\mathbb{T}:G\times\X\rightarrow\X$, see \cref{fig:transforms} for examples of $\mathcal{T}$ and $\mathcal{D}$.

\begin{figure}[tb]
  \hspace{-0.4em}\includegraphics[width=0.57\textwidth]{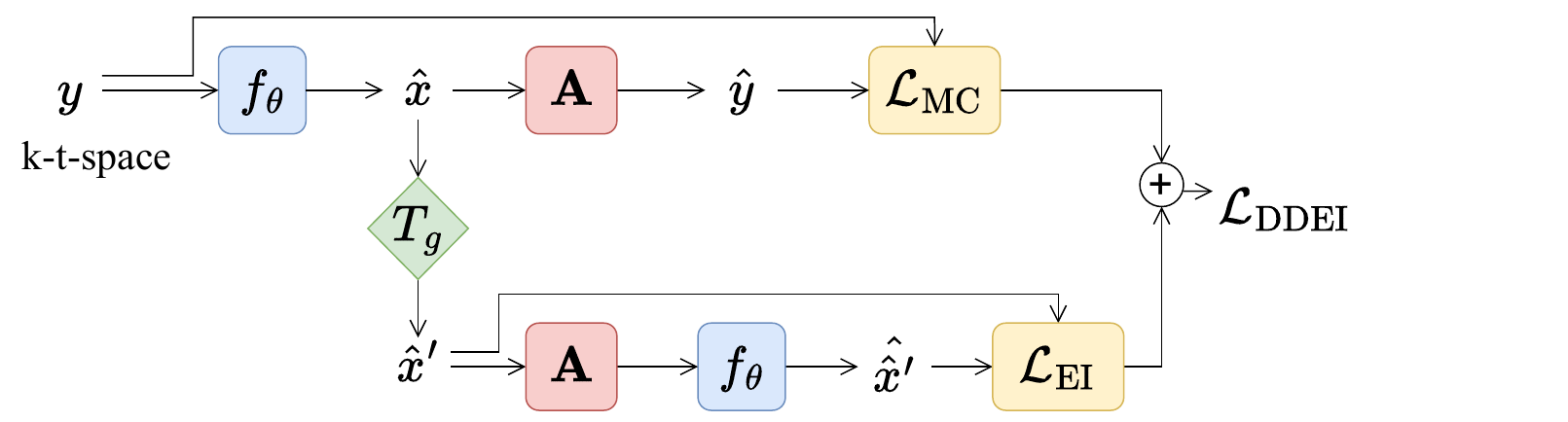}
  \caption{Our DDEI ground truth-free loss function. $\A$ is an undersampled Fourier operator, $\ft$ is any neural net backbone, and $\Tg$ is a random diffeo-temporal transform (\cref{fig:transforms}). At inference, we simply compute $\xhat=\ft(\y)$.}
  \label{fig:framework}
\end{figure}

During training, we draw $g\in G$ and the nonlinear action $\mathbb{T}(g,x)=\Tg(x)$. Then $\y=\A\x=\A\cdot\Tgi\cdot\Tg(\x)=\A_g(\x^{\prime})$ where $\A_g=\A\cdot\Tgi$ is now a nonlinear mapping with $\x^{\prime}=\Tg(\x)\in\X$. Thus the simple assumption that the signal set $\X$ is $G$-invariant allows us to image $\X$ via multiple transformed nonlinear operators $\A_g(\cdot)$ and ``see into" the nullspace. From \cite{chen_equivariant_2021}, this leads to $G$-equivariance of the imaging system. We can constrain this by constructing the \textbf{DDEI} GT-free loss inspired by \cite{chen_equivariant_2021}, which incorporates the non-linear diffeo-temporal group action $\Tg$:

\begin{equation}
\L_\text{DDEI}(\theta;\y)=\L_\text{MC}+\alpha\L(\Tg\cdot \ft(\y),\ft(\A\cdot\Tg\cdot \ft(\y)))
    \label{eq:loss_function}
\end{equation}

\noindent where $\alpha$ is a constant that we set to 1, and $\mathcal{L}_\text{MC}$ is the measurement consistency loss. See \cref{fig:framework} for a graphical overview.

\textbf{Noisy adaptation} In the presence of Gaussian noise $\mathbf{\epsilon}$ on the measurements, we replace $\L_\text{MC}$ with the SURE loss from \cite{chen_robust_2022} which is an unbiased estimator of the supervised MSE.

\section{Related work}

We consider other fully unsupervised DL methods that attempt to learn from measurements alone, so we discount pretrained diffusion methods, paired measurements, and classical CS methods. Note that our work is distinct to motion-correction methods: we reconstruct full image sequences.

\textbf{Measurement splitting} A popular set of state-of-the-art approaches includes SSDU \cite{acar_self-supervised_2021}, Phase2Phase \cite{eldeniz_phase2phase_2021} and Artifact2Artifact (used in \cite{gan_deformation-compensated_2022}). They randomly split (in various ways) the k-t-space into two sets at each iteration. One is used as the input and the other to construct the loss at the output:

\begin{equation}
    \L_\text{split}=\L(\M_2\A \ft(\M_1\y),\M_2\y),\; \xhat=\ft(\y,\A)
    \label{eq:splitting}
\end{equation}

\textbf{Equivariant imaging} EI \cite{chen_equivariant_2021} is a state-of-the-art method that constrains the set of solutions using group invariance, currently using a limited set of spatial invariances such as rotation $\mathcal{R}=\text{SO}(2)$, applied to static MRI in \cite{chen_robust_2022}, and homography \cite{wang_perspective-equivariance_2024}. Here we extend EI using temporal invariances and further geometric invariance using diffeomorphisms.

\textbf{Other unsupervised} Several methods perform test-time optimisation on an untrained NN, inspired by Deep Image Prior and Implicit Neural Representations \cite{sultan_deep_2024,kunz_implicit_2024}. However, these result in lengthy inference, which is undesirable in practice, and cannot take advantage of large training datasets. Deep Restoration Prior \cite{hu_restoration_2023} performs optimisation on top of a well-trained model so can be used alongside our work. We therefore do not compare our method to these.

\begin{figure*}[tb]
  \centering
  \includegraphics[width=1.0\textwidth]{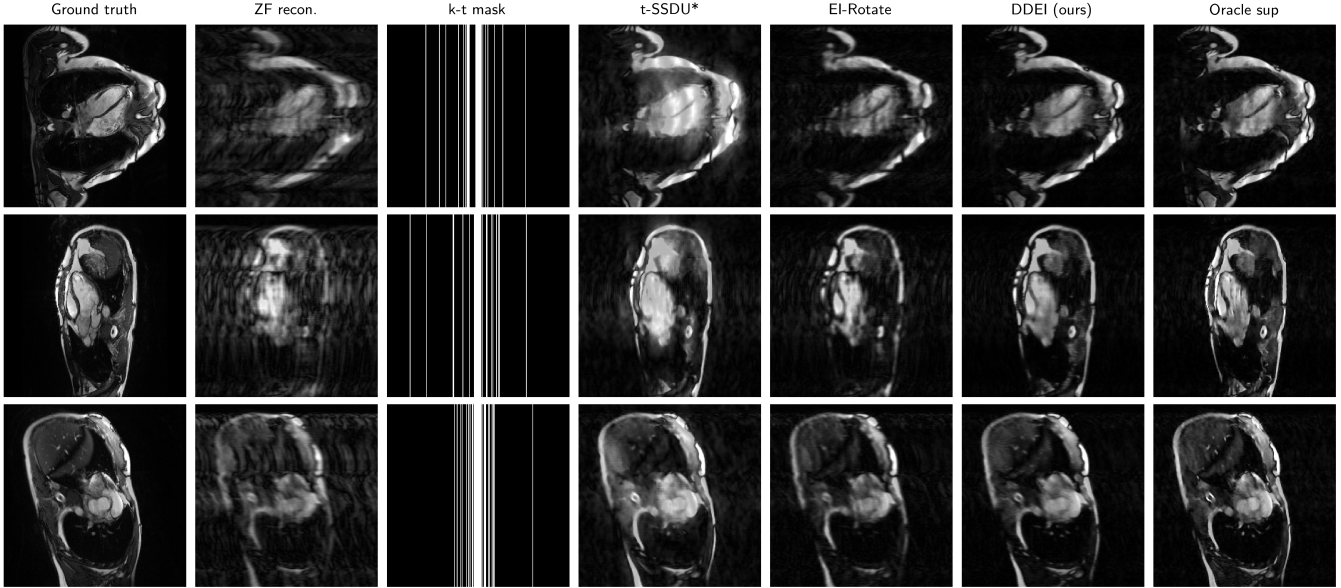}
  \caption{Test set example cardiac long axis views (above 2 rows) and short axis slice (below) reconstruction results on 8x accelerated dynamic MRI at time-step $t=0$. For full videos see code link\protect\footnotemark[1]. Best performing methods are compared.}
  \label{fig:test_acc_8}
\end{figure*}

\section{Experiment and result}

\textbf{Dataset} We demonstrate our method on a real-world cardiac cine MRI dataset from the 2023 CMRxRecon challenge \cite{lyu_state---art_2024} training set, so that we have gated GT to evaluate our method.  The gatedness of the dataset does not affect comparisons between unsupervised methods, as they do not assume periodicity. We emphasise we do not require GT during training and our method can equally be applied to ungated in-vivo k-t-space measurements. The data consists of 120 patients (split 80\% train-test), each providing 1 short-axis slice, to speed up training, and at most 3 long-axis views (2, 3, 4 chamber) to a total of 417 2D+t sequences. The data was provided as fully-sampled single-coil cine k-t-space sequences, with each cropped to $512\times256\times12$ frames. Pseudo-GT is obtained by inverse FFT and standardisation following fastMRI \cite{zbontar_fastmri_2018}. Then, we simulate an undersampled 2D+t measurement operator using random Cartesian Gaussian time-i.i.d masking, at 8x acceleration (acc.) with fixed central ACS lines. 

\textbf{Implementation details} We implement our method and competitors in our code \footnote{https://github.com/Andrewwango/ddei} using the DeepInverse \footnote{https://deepinv.github.io} library. For our loss, since diffeomorphisms are very general, we relax the constraint and enforce approximate equivariance by taking a subset of diffeomorphisms $\mathcal{D}^\prime$ as defined by the continuous piecewise-affine-based transforms (CPAB) \cite{freifeld_transformations_2017} for small distortions (see \cref{fig:transforms}). These are equivalent to a subset that lie on a Riemannian manifold and also benefit from an efficient GPU implementation. We also reconsider the 2D rotation subgroup from \cite{chen_equivariant_2021} $\mathcal{R}=\text{SO}(2)$ such that $\mathcal{D}=\mathcal{D}^\prime\bigotimes\mathcal{R}$.

For $\ft$ we use a very small convolutional recurrent neural network \cite{qin_convolutional_2019}, a lightweight unrolled network with 2 unrolled iterations and 1154 parameters. We emphasise that our framework is NN-agnostic and any state-of-the-art NN can be used as the backbone. We report metrics as defined in \cite{zbontar_fastmri_2018}.

\textbf{Baselines} We compare DDEI to unsupervised baselines: zero-filled (ZF) reconstruction using the inverse FFT $\xhat={\A}^H \y$, measurement consistency (MC) $\L_\text{MC}$, and popular existing methods: the measurement splitting method \textit{t-SSDU} \cite{acar_self-supervised_2021} (i.e. Artifact2Artifact \cite{gan_deformation-compensated_2022}) with 60\% Gaussian masking, \textit{t-SSDU*} which is a modification of t-SSDU that performs $\xhat=\frac{1}{N}\Sigma_{i=1}^n \ft(\M_i\y,\M_i\A)$ at test-time as inspired by \cite{hendriksen_noise2inverse_2020}, Phase2Phase \cite{eldeniz_phase2phase_2021}, and standard EI with rotation only from \cite{chen_equivariant_2021}.

\textbf{Results} Quantitative results from 8x acc. are shown in \cref{tab:test_acc_8} and sample reconstruction frames are shown in \cref{fig:test_acc_8}. Our method, DDEI, improves on existing methods by a significant margin, and approaches ``oracle" supervised performance $\mathcal{L}_\text{sup}$. Its reconstructions are much cleaner with fewer artifacts, sharper edges, and smoother in time, due to our joint spatial diffeo-temporal equivariance. We vastly improve on all methods involving splitting including SSDU \cite{acar_self-supervised_2021} and Phase2Phase \cite{eldeniz_phase2phase_2021}, showing that splitting cannot remain competitive in higher undersampling. Our method also generalises across per-sample masks and multiple cardiac views. 

Note that, since we are comparing frameworks and not the NN backbones, the methods can easily be scaled and improved by using larger NN architectures such as transformers. 

We also report results in the Gaussian noise ($\sigma=3$) scenario in \cref{tab:test_noisy}, and see that simply adding SURE to DDEI denoises the input resulting in a robust performance without GT that remains competitive with supervised learning.

\begin{table}[tb]
\centering
\begin{tabular}{llll}
\hline
Loss & PSNR $\uparrow$ & SSIM $\uparrow$ & NMSE $\downarrow$ \\ \hline
ZF & 27.1\textpm 1.6 & .686\textpm .047 & .366\textpm .064 \\
MC & 27.1\textpm 1.6 & .686\textpm .047 & .366\textpm .064 \\
t-SSDU \cite{acar_self-supervised_2021}  & 17.1\textpm2.3 & .520\textpm.053 & 3.89\textpm1.26 \\
t-SSDU* & 29.4\textpm 2.5 & .710\textpm .059 & .235\textpm .100 \\
Phase2Phase \cite{eldeniz_phase2phase_2021} & 27.4\textpm 2.1 & .768\textpm .045 & .368\textpm .150 \\
EI-$\mathcal{R}$otate \cite{chen_equivariant_2021} & 30.1\textpm 1.5 & .798\textpm .039 & .193\textpm .063 \\
\color{blue}\textbf{DDEI (ours)} & \color{blue}\textbf{33.3}\textpm 1.7 & \color{blue}\textbf{.864}\textpm .034 & \color{blue}\textbf{.096}\textpm .043 \\
(Oracle supervised) & 36.4\textpm1.4 & .887\textpm.029 & .045\textpm.015 \\ \hline
EI-$\mathcal{R}$-$\mathcal{T}$ (Ablation) & 33.0\textpm 1.7 & .860\textpm .034 & .102\textpm .046 \\ \hline
\end{tabular}
\caption{Test set results for 8x accelerated dynamic MRI. Best unsupervised method in \textbf{bold}.}
\label{tab:test_acc_8}
\end{table}

\begin{table}[tb]
\centering
\begin{tabular}{llll}
\hline
Loss & PSNR $\uparrow$ & SSIM $\uparrow$ & NMSE $\downarrow$ \\ \hline
ZF & 22.9\textpm 1.6 & .141\textpm .034 & .957\textpm .069 \\
t-SSDU* & 24.0\textpm 2.1 & .205\textpm .043 & .767\textpm .248 \\
DDEI (ours) & 26.8\textpm 1.5 & .274\textpm .046 & .392\textpm .048 \\
\color{blue}\textbf{DDEI-SURE} (ours) & \color{blue}\textbf{30.7}\textpm 1.5 & \color{blue}\textbf{.817}\textpm .035 & \color{blue}\textbf{.168}\textpm .056 \\
(Oracle supervised) & 33.6\textpm 1.5 & .845\textpm .032 & .086\textpm .026 \\ \hline
\end{tabular}
\caption{Test set results for noisy 8x acc. dynamic MRI.}
\label{tab:test_noisy}
\end{table}

\section{Conclusion}

In this paper we propose DDEI, a simple unsupervised framework for practical future MRI systems to reconstruct high-quality true body motion sequences, where ground truth is impossible to obtain. This is crucial as supervised methods require gated pseudo-GT, which can never learn true motion. DDEI learns from undersampled measurements alone using a NN-agnostic loss, by constraining the signal set with temporal and diffeomorphic group invariance, to recover more information lost in the undersampling. We show that, on accelerated cardiac imaging, DDEI approaches supervised learning and highly outperforms existing methods.

In future, since using simulated measurements from pseudo-GT are not true k-t-space, we should train from raw k-t-space data and use radiologist scoring as evaluation.

\textbf{Acknowledgments} This study was conducted retrospectively using human subject data made available in open access by \cite{lyu_state---art_2024}. Ethical approval was not required as confirmed by the license
attached with the open access data. The authors have no relevant financial or non-financial interests to disclose.

\bibliographystyle{IEEEbib}
\bibliography{references}

\begin{thebibliography}{10}

\bibitem{lyu_state---art_2024}
J.~Lyu, C.~Qin et~al.,
\newblock ``The state-of-the-art in {Cardiac} {MRI} {Reconstruction}: {Results} of the {CMRxRecon} {Challenge} in {MICCAI} 2023,'' Apr. 2024.

\bibitem{liu_rare_2020}
J.~Liu, Y.~Sun et~al.,
\newblock ``{RARE}: {Image} {Reconstruction} {Using} {Deep} {Priors} {Learned} {Without} {Groundtruth},''
\newblock {\em IEEE Journal of Selected Topics in Signal Processing}, vol. 14, no. 6, pp. 1088--1099, Oct. 2020.

\bibitem{lim_3d_2019}
Y.~Lim, Y.~Zhu et~al.,
\newblock ``{3D} dynamic {MRI} of the vocal tract during natural speech,''
\newblock {\em Magnetic Resonance in Medicine}, vol. 81, no. 3, pp. 1511--1520, 2019.

\bibitem{shimron_implicit_2022}
E.~Shimron, J.~I. Tamir et~al.,
\newblock ``Implicit data crimes: {Machine} learning bias arising from misuse of public data,''
\newblock {\em Proceedings of the National Academy of Sciences}, vol. 119, no. 13, Mar. 2022.

\bibitem{gan_deformation-compensated_2022}
W.~Gan, Y.~Sun et~al.,
\newblock ``Deformation-{Compensated} {Learning} for {Image} {Reconstruction} {Without} {Ground} {Truth},''
\newblock {\em IEEE Transactions on Medical Imaging}, vol. 41, no. 9, pp. 2371--2384, Sept. 2022.

\bibitem{chen_equivariant_2021}
D.~Chen, J.~Tachella et~al.,
\newblock ``Equivariant {Imaging}: {Learning} {Beyond} the {Range} {Space},''
\newblock in {\em 2021 {IEEE}/{CVF} {International} {Conference} on {Computer} {Vision} ({ICCV})}, Oct. 2021.

\bibitem{chen_robust_2022}
D.~Chen, J.~Tachella et~al.,
\newblock ``Robust {Equivariant} {Imaging}: a fully unsupervised framework for learning to image from noisy and partial measurements,''
\newblock in {\em 2022 {IEEE}/{CVF} {Conference} on {Computer} {Vision} and {Pattern} {Recognition} ({CVPR})}, June 2022.

\bibitem{acar_self-supervised_2021}
M.~Acar, T.~Çukur et~al.,
\newblock ``Self-supervised {Dynamic} {MRI} {Reconstruction},''
\newblock in {\em Machine {Learning} for {Medical} {Image} {Reconstruction}}. 2021, pp. 35--44, Springer International Publishing.

\bibitem{eldeniz_phase2phase_2021}
C.~Eldeniz, W.~Gan et~al.,
\newblock ``{Phase2Phase}: {Respiratory} {Motion}-{Resolved} {Reconstruction} of {Free}-{Breathing} {Magnetic} {Resonance} {Imaging} {Using} {Deep} {Learning} {Without} a {Ground} {Truth} for {Improved} {Liver} {Imaging},''
\newblock {\em Investigative Radiology}, vol. 56, no. 12, Dec. 2021.

\bibitem{wang_perspective-equivariance_2024}
A.~Wang and M.~Davies,
\newblock ``Perspective-{Equivariance} for {Unsupervised} {Imaging} with {Camera} {Geometry},'' Mar. 2024,
\newblock arXiv:2403.09327 [cs, eess].

\bibitem{sultan_deep_2024}
M.~A. Sultan, C.~Chen et~al.,
\newblock ``Deep {Image} prior with {StruCtUred} {Sparsity} ({DISCUS}) for dynamic {MRI} reconstruction,'' May 2024,
\newblock arXiv:2312.00953.

\bibitem{kunz_implicit_2024}
J.~F. Kunz, S.~Ruschke et~al.,
\newblock ``Implicit {Neural} {Networks} with {Fourier}-{Feature} {Inputs} for {Free}-breathing {Cardiac} {MRI} {Reconstruction},'' Jan. 2024,
\newblock arXiv:2305.06822.

\bibitem{hu_restoration_2023}
Y.~Hu, M.~Delbracio et~al.,
\newblock ``A {Restoration} {Network} as an {Implicit} {Prior},'' Oct. 2023,
\newblock arXiv:2310.01391 [eess].

\bibitem{zbontar_fastmri_2018}
J.~Zbontar, F.~Knoll et~al.,
\newblock ``{fastMRI}: {An} {Open} {Dataset} and {Benchmarks} for {Accelerated} {MRI},'' Nov. 2018.

\bibitem{freifeld_transformations_2017}
O.~Freifeld, S.~Hauberg et~al.,
\newblock ``Transformations {Based} on {Continuous} {Piecewise}-{Affine} {Velocity} {Fields},''
\newblock {\em IEEE Transactions on Pattern Analysis and Machine Intelligence}, vol. 39, no. 12, pp. 2496--2509, Dec. 2017.

\bibitem{qin_convolutional_2019}
C.~Qin, J.~Schlemper et~al.,
\newblock ``Convolutional {Recurrent} {Neural} {Networks} for {Dynamic} {MR} {Image} {Reconstruction},''
\newblock {\em IEEE Transactions on Medical Imaging}, vol. 38, no. 1, pp. 280--290, Jan. 2019.

\bibitem{hendriksen_noise2inverse_2020}
A.~A. Hendriksen, D.~M. Pelt et~al.,
\newblock ``{Noise2Inverse}: {Self}-{Supervised} {Deep} {Convolutional} {Denoising} for {Tomography},''
\newblock {\em IEEE Transactions on Computational Imaging}, vol. 6, pp. 1320--1335, 2020.

\end{thebibliography}

\end{document}